\documentclass[letterpaper]{article}
\usepackage{aaai}
\usepackage{times}
\usepackage{helvet}
\usepackage{courier}
\usepackage{graphicx} 
\usepackage{multirow} 
\usepackage{url}
\usepackage{float}
\usepackage{textcomp}
\begin{document}
%
\title{AbAffinity: A Large Language Model for Predicting Antibody Binding Affinity against SARS-CoV-2}
\author{Faisal Bin Ashraf$^1$ \quad Animesh Ray$^2$ \quad Stefano Lonardi$^1$\\
$^1${Department of Computer Science and Engineering,University of California, Riverside, 92521, CA, United States}  \\ 
$^2${Riggs School of Applied Life Sciences, Keck Graduate Institute, Claremont, 91711, CA, United States}
}

\maketitle
\footnotetext{Presented at the FMs4Bio Workshop at AAAI 2025.}
\begin{abstract}
\begin{quote}
Machine learning-based antibody design is emerging as one of the most promising approaches to combat infectious diseases, due to significant advancements in the field of artificial intelligence and an exponential surge in experimental antibody data (in particular related to COVID-19). The ability of an antibody to bind to an antigens (called \emph{binding affinity}) is one of the the most critical properties in designing neutralizing antibodies. In this study we introduce Ab-Affinity, a new large language model that can accurately predict the binding affinity of antibodies against a target peptide, e.g., the SARS-CoV-2 spike protein. Code and model are available at \url{https://github.com/ucrbioinfo/AbAffinity}. 
\end{quote}
\end{abstract}

\section{Introduction}

Antibodies are proteins composed of two identical polypeptide chains, termed ``heavy chains" and two identical ``light chains", respectively, connected by disulfide bonds. Each light chain consists of one variable domain and one constant domain, while each heavy chain contains one variable domain and 3-4 constant domains. Each antibody possesses an antigen binding site, called \emph{paratope}, typically buried within the variable domain's complementarity-determining regions (CDRs). The paratope recognizes and binds specifically to a target region on the surface of an antigen, known as the \emph{epitope}.   CDRs together constitute the binding site of the antibody, and their sequence diversity allows antibodies to recognize a wide range of antigens with high specificity. The binding interaction between the antibody and antigen is highly specific. This binding may trigger various immune responses, including the neutralization of a live virus, a toxin or an antigen with catalytic activities, leading to the eventual elimination of the antigen from the organism. 

\begin{figure}[t]
    \centering
    \includegraphics[width=\linewidth]{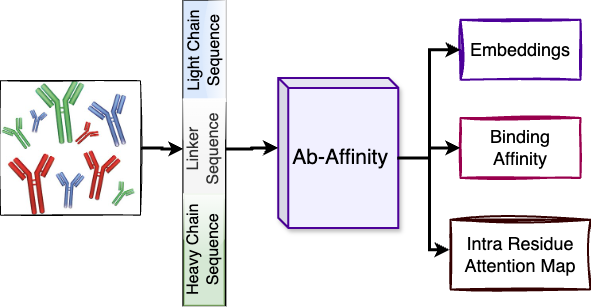}
    \caption{Overview of Ab-Affinity. Ab-Affinity predicts the binding affinity against a specific target peptide; it can also provide residue-residue contact maps and an embedding of the input sequence. }
    \label{fig:overview}
\end{figure}

Measuring the binding affinity of an antibody against an antigen is critical for understanding immune responses elicited by vaccines and infections, as well as for designing antibodies with therapeutic applications. Such measurements often involve biophysical and biochemical techniques such as Surface Plasmon Resonance (SPR), Enzyme-Linked Immunosorbent Assay (ELISA), or Bio-Layer Interferometry (BLI), which generally includes immunization of animals to obtain a host of candidate antibodies, their purification, and testing of the candidates for choosing the optimum molecules that exhibit high-specificity and sufficient strength of epitope-paratope interaction. These approaches incur significant time and cost. Computational approaches hold promise to replace some of these time-intensive steps to narrow down the list of candidate antibodies for final testing.

Predicting the specificity and strength of antigen-antibody interaction for protein antigens is a sub-problem of predicting specific protein-protein interactions. However, the former problem is harder because the paratope and the epitope regions of many antibodies and their cognate antigens are often flexible, and contain relatively mobile intrinsically disordered protein regions (IDPRs) when not in complex with the cognate antigen \cite{uversky2021mobility,macraild2016antibody}. Structures of flexible and mobile regions of proteins are relatively poorly represented in protein structure databases because they are technically harder to determine; training data for learning structural rules are consequently sparse. Predicting these interactions from sequences can be advantageous if the structural and functional roles of amino acids in the sequence are effectively identified and modeled.

The problem of predicting binding affinity has been extensively studied in the literature. For instance, ISLAND \cite{abbasi2020island} uses a comprehensive feature extraction from protein sequence combined with regression but achieves a modest predictive accuracy. DeepDTA \cite{ozturk2018deepdta} improves upon ISLAND by relying on Convolutional Neural Networks (CNN) for protein-ligand interactions. More recent models like DG-Affinity \cite{yuan2023dg} and CSM-AB \cite{myung2022csm} integrate large language models and graph-based techniques, thus enhancing the accuracy of predictions. Tag-LLM \cite{shen2024tag} and FAbCon \cite{barton2024generative} further refine these capabilities by using task-specific tags and fine-tuned large language models for antibody-antigen interactions.  The use of pre-trained protein language models in DG-Affinity \cite{yuan2023dg} has shown superior performance in predicting antibody binding affinities compared to many other methods. Large language models were also used to predict the binding affinity of antibodies against the SARS-CoV-2 spike protein \cite{li2023machine}. However, due to the specificity of antibodies to antigens, existing multi-target affinity prediction models are unsuitable for SARS-CoV-2. Therefore, an efficient binding affinity prediction model is essential targeting this virus.  

In this study, we introduce a large language model called Ab-Affinity that was trained on single-chain fragment variable (scFv) sequences from multiple libraries of engineered antibodies.  Ab-Affinity provides meaningful representation (i.e., embedding) of antibodies which align with binding properties and thermostability proprieties against SARS-CoV-2, and can be used to create the binding landscape of antibodies. Second, Ab-Affinity provides intra-residue attention maps of antibodies which can be used to explain differences of amino-acid interactions for strong and weak binding. Figure~\ref{fig:overview} illustrates the overall framework of this study. 

\section{Methods}


\subsection{Dataset}
\label{section:dataset}
\begin{figure*}[t]
    \centering
    \includegraphics[width=0.75\linewidth]{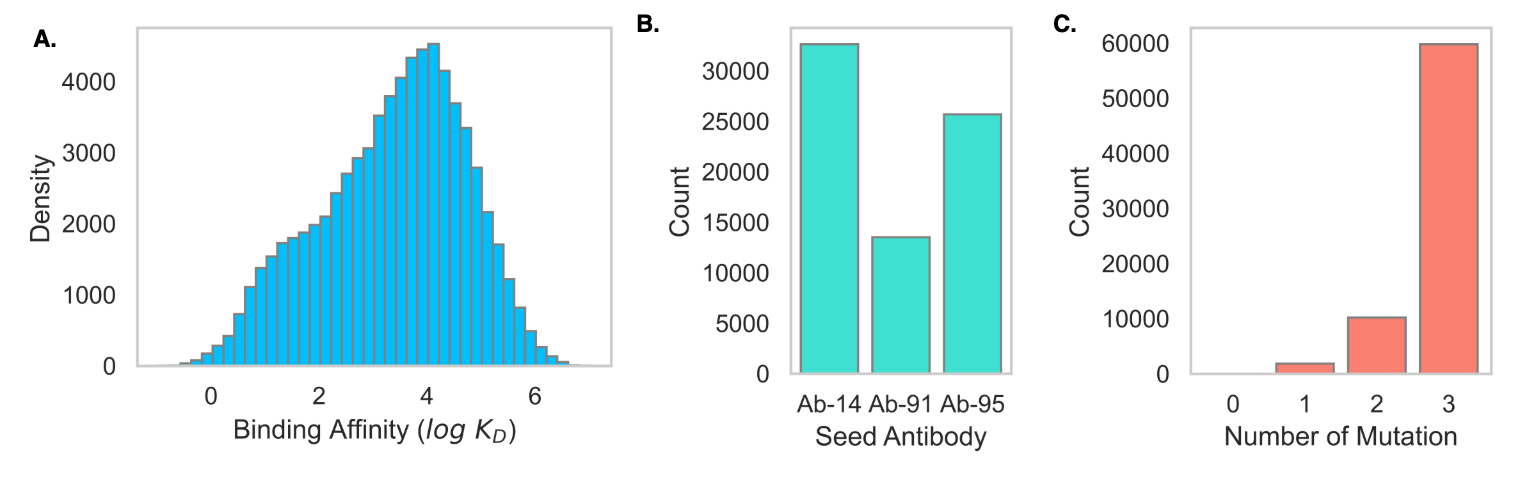}
    \caption{Dataset description. a. Distribution of Binding Affinity ($\log~K_D$); b. Distribution of antibodies from each seed; c. Distribution of antibodies with mutation.}
    \label{fig:dataset} 
\end{figure*}

The training used a dataset of single-chain fragment variable (scFv) antibody sequences with the associated binding scores against a peptide in the SARS-CoV-2 HR2 region \cite{engelhart2022dataset}. The SARS-CoV-2 spike protein has mutated, forming variants such as Wuhan, Alpha, Delta, Omicron, and others. The selected HR2 peptide, conserved across all variants of SARS and MERS spike proteins, is crucial for evaluating antibodies against the broader coronavirus group. The dataset was obtained by introducing one, two, and three amino acid changes into the sequences of three candidate antibodies obtained from a phage display library that bound to the HR2 antigen polypeptide: Ab-14-VH and Ab-14-VL, Ab-91-VH, Ab-95-VH and Ab-95-VL. The binding affinity ($K_D$) of each of the 104,972 resulting variants the selected target peptide was estimated in three independent biological replicates \cite{engelhart2022dataset}. Each $K_D$ value was the equilibrium dissociation constant estimated by an indirect competitive binding assay. The experimental assay was conducted in triplicate for each antibody, and the full dataset contained three $K_D$ values for each antigen-antibody pair. We preprocessed the interaction data by taking the arithmetic mean of the two closest $K_D$ values for each antigen-antibody interaction and by eliminating the third value to reduce the outlier effect. Antibodies that had missing values across all three replicates were disregarded from the analysis. After preprocessing, a total of 71,834 unique antibodies were utilized for model training. Ab-Affinity was trained on all antibody sequences (from each seed antibody) and their corresponding log-transformed binding affinity $log_{10}(K_d)$. Distributions of training antibodies are shown in Figure~\ref{fig:dataset}. 

\subsection{Model Architecture}

\begin{figure*}[ht]
    \centering
    \includegraphics[width=0.8\linewidth]{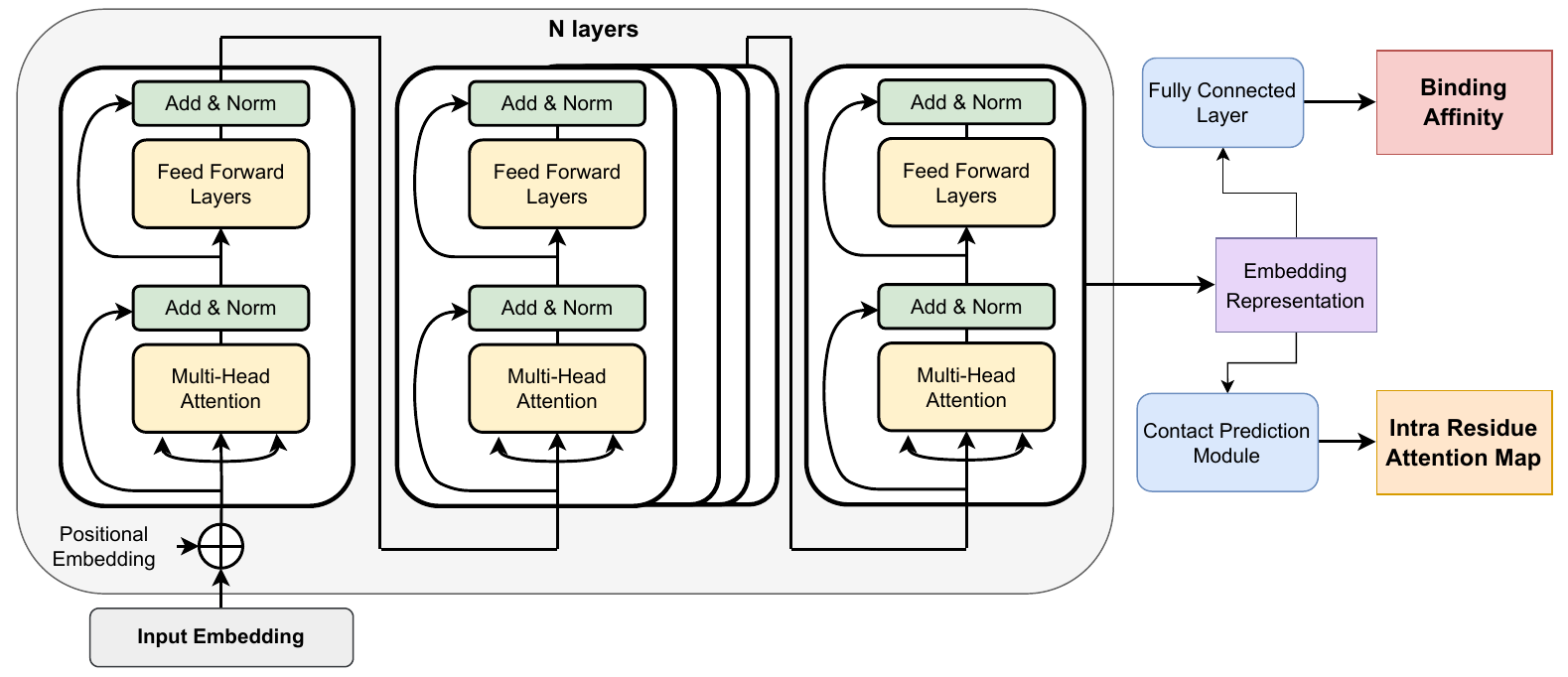}
    \caption{Model Architecture of Ab-Affinity}
    \label{fig:model_architecture} 
\end{figure*}

Ab-Affinity's architecture is based on BERT \cite{devlin2018bert}, as implemented in ESM-2 \cite{lin2023evolutionary} (see Figure~\ref{fig:model_architecture}). We chose the BERT architecture because the dataset consists of amino acid sequences with a few point mutations. BERT effectively captures long-range dependencies, providing meaningful representations that align with binding affinity changes caused by these mutations.  Ab-Affinity contains $N$ sequential layers of encoder blocks. Each encoder block consists of multi-head attention layers \cite{vaswani2017attention} followed by feed-forward layers. The value of $N$ determines the model size. We have tested $N = 6, 12,$ and $33$, which resulted in a model with $8M, 35M,$ and $650M$ parameters, respectively. We chose these values for  N  based on the ESM-2 study \cite{lin2023evolutionary}, which demonstrated the impact of model size on the performance of this BERT-based architecture. We used the last encoder layer output as the sequence representation (i.e., the \emph{embedding}). Depending on the choice of $N$ the embedding of the sequence had $320, 480$ and $1280$ dimensions, respectively. We added one fully connected layer to predict the binding affinity from the embedding of the sequence.  

\subsection{Training}
We used Mean Squared Error (MSE) as the loss function, and Adam optimizer to optimize the parameters. We used 85\% of the data to train the model (maintaining the distribution of affinity values). The remaining 15\% was used to validate the model. We trained our model using four cores NVIDIA A100 (80GB) with a batch size of 128 for 100 epochs. For the amino-acid contact analysis, we used the method by Rao \emph{et. al.} \cite{rao2020transformer}. We fine-tuned the pre-trained ESM-2 protein language model encoder layers, leveraging the knowledge acquired from the entire protein database to train our model. Additionally, we trained a model with randomly initiated weights to understand the impact of pretrained protein knowledge on binding affinity prediction.  We saved the best-performing model based on the Pearson correlation coefficient on the validation set for each setup of training. Ab-Affinity, with 33 layers and fine-tuned from ESM-2, is the best-performing model among all the trained models.

\subsection{t-SNE visualization of embeddings}

The T-Distributed Stochastic Neighbor Embedding (t-SNE) \cite{van2008visualizing} was used to reduce high dimensional sequence representation to 2-D space to plot the antibodies. We carried out the dimensionality reduction using Python scikit-learn package and set the perplexity at 200.

\section{Experimental Results}


\subsection{Ab-Affinity Predicts the Binding Affinity of Antibodies Against SARS-CoV-2} 


We first investigated whether Ab-Affinity creates meaningful latent space features (i.e., embeddings) for the antibodies after training. To this end, we visualized Ab-Affinity's embeddings in a 2D space using t-distributed Stochastic Neighbor Embedding (t-SNE) maps. To compare the performance of our methods with the best existing method, we also visualized the embeddings produced by ESM-2.

\begin{figure}[ht]
    \centering
    \includegraphics[width=0.99\linewidth]{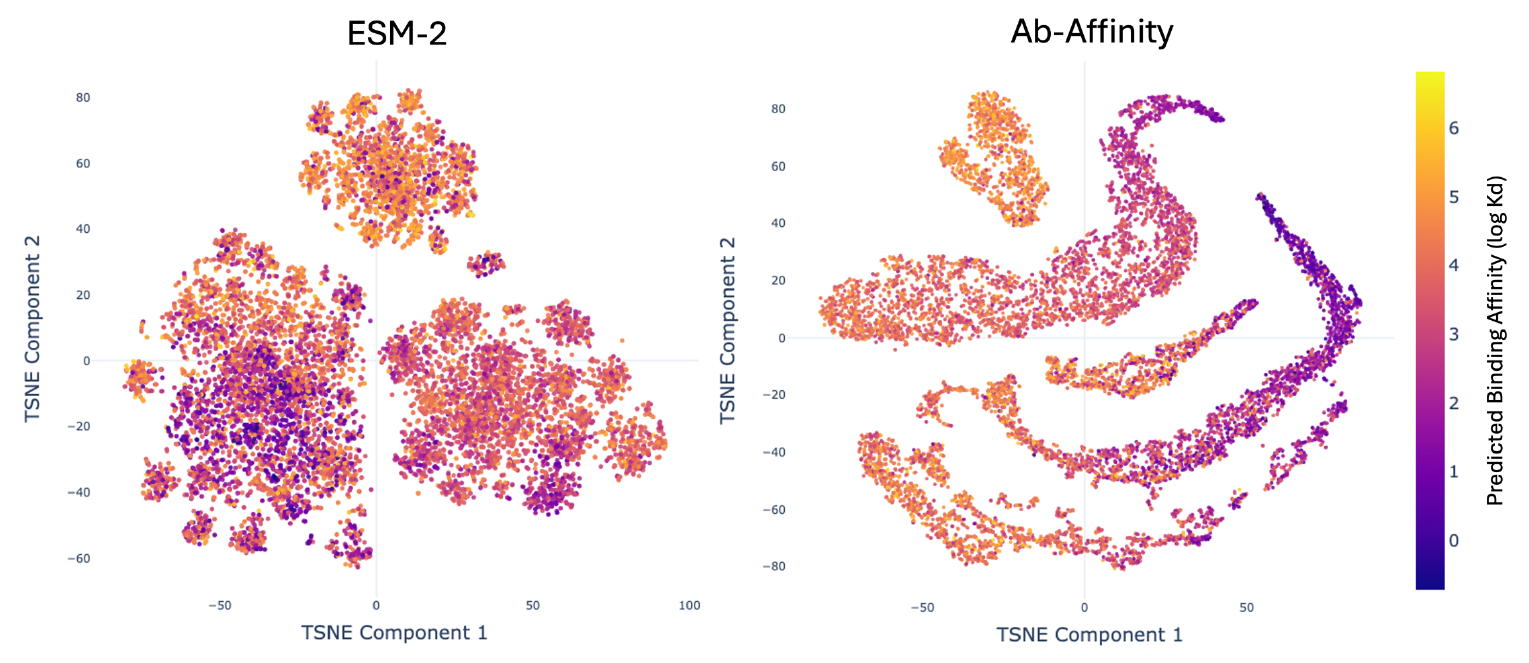}
    \caption{t-SNE representation of the embedding produced by ESM-2 and Ab-Affinity; antibodies are colored according to their predicted binding affinity.}
    \label{fig:emb_2D}
\end{figure}

In the t-SNE plots in Figure~\ref{fig:emb_2D} each point is an antibody and the color illustrates its corresponding binding affinity. Observe in Figure~\ref{fig:emb_2D} that the embeddings produced by Ab-Affinity display a smooth gradient of binding affinity, i.e., the value of $\log K_d$ monotonically decreases. By contrast, the ESM-2 embeddings in Figure~\ref{fig:emb_2D} do not show a clear gradient descent along the t-SNE components.

\begin{figure*}[bht]
    \centering
    \includegraphics[width=0.97\textwidth]{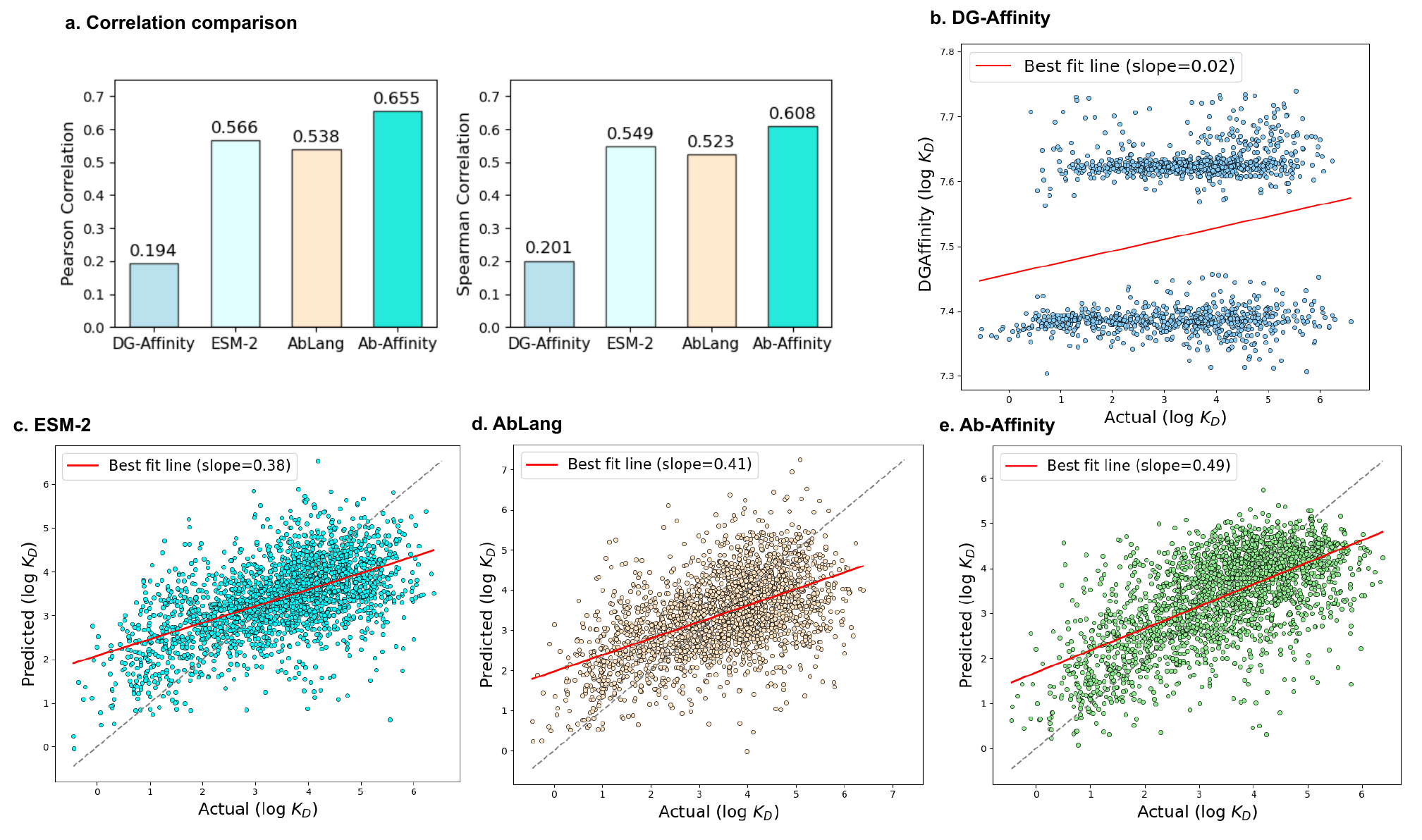} 
    \caption{Comparing affinity prediction models: (a) Pearson and Spearman correlation for DG-Affinity \small{(p-values = $3.86\times10^{-14}$, and $3.38\times10^{-15}$)}, ESM-2 \small{(p-values = $8.03\times10^{-198}$, and $8.02\times10^{-198}$)}, AbLang \small{(p-values = $1.24\times10^{-175}$, and $1.02\times10^{-163}$)} and Ab-Affinity \small{(p-values = $4.03\times10^{-261}$, and $8.65\times10^{-217}$)}; scatter plot for actual vs. predicted binding affinity for (b) DG-Affinity, (c) ESM-2, (d) AbLang, (e) Ab-Affinity (includes antibodies for all three seeds)}
    \label{fig:aff_comp}
\end{figure*}

\begin{table*}[h]
\centering
\caption{Comparing Ab-Affinity against other binding affinity predictors on 14H and 14L dataset }
\begin{tabular}{ll|cc|cc}
\hline\hline
\multicolumn{1}{c}{\multirow{2}{*}{\textbf{Model}}} & \multicolumn{1}{c|}{\multirow{2}{*}{\textbf{Ref}}} & \multicolumn{2}{c|}{\textbf{14H}} & \multicolumn{2}{c}{\textbf{14L}} \\ \cline{3-6} 
\multicolumn{2}{c|}{} & Pearson & Spearman & Pearson & Spearman \\ \cline{1-6}
Ens-Grad & \cite{liu2020antibody} & 0.601 & 0.476 & 0.637 & 0.645 \\
ESM-F & \cite{he2024novo} & 0.634 & 0.516 & 0.674 & 0.681 \\
AntiBERTa2 & \cite{barton2024enhancing} & 0.623 & 0.545 & 0.673 & 0.684 \\
AbMAP & \cite{singh2023learning} & 0.606 & 0.510 & 0.674 & 0.685 \\
A2Binder & \cite{he2024novo} & 0.642 & \textbf{0.553} & 0.683 & 0.688 \\
Ensembles-14H & \cite{li2023machine} & N/A & 0.512 & N/A & N/A \\
Ensembles-14L & \cite{li2023machine} & N/A & N/A & N/A & 0.688 \\
\textbf{Ab-Affinity} & [this] & \textbf{0.652} & 0.526 & \textbf{0.712} & \textbf{0.713} \\ \hline\hline 
\end{tabular}

\label{tab:aff_comparison2}
\end{table*}

To evaluate the performance of our method, we then compared Ab-Affinity’s ability to predict binding affinity with those of three other LLM-based methods, namely DG-Affinity, ESM-2, and AbLang. The same test dataset, which was held out from training and not utilized during model development, was used to predict binding affinity and compare the performance of the methods.  To proceed with this comparison, We recall that DG-Affinity uses a pre-trained language model combined with a ConvNext-based architecture for prediction, which outperforms 26 other methods on an independent antibody dataset \cite{yuan2023dg}.  ESM-2 was pre-trained on the UniRef50 protein database and generated 640-dimensional antibody sequence embeddings, which we fed into a simple linear regression model. AbLang uses separate embeddings for the heavy and light chains of antibodies, which we concatenated into a 1534-dimensional vector and used in a linear regression model to predict the affinity score. 

Figure~\ref{fig:aff_comp} illustrates (a) Pearson and Spearman-rank correlation coefficients for the four methods on the test set, (b-e) scatter plots for actual vs. predicted affinity for the four methods. The highest Pearson/Spearman correlation coefficients were observed with the predictions from Ab-Affinity. Observe that DG-Affinity has the lowest Pearson correlation coefficient of 0.194, presumably because although it used an LLM pre-trained on antibodies, the architecture head itself was not trained with any SARS-CoV-2-specific data. Surprisingly, ESM-2, which was pre-trained on ``generic'' proteins, produced a good Pearson/Spearman correlation. AbLang, which is instead trained on antibody sequences, also achieved a good Pearson/Spearman correlation. The scatter plots of actual vs. predicted binding affinity in Figure~\ref{fig:aff_comp}(b-e) show that the best-fit lines for Ab-Affinity have the highest slope, indicating the best predictive accuracy.

To explore whether the prediction methods are sensitive to the choice of the test dataset, we compared Ab-Affinity with the other three methods using a different dataset. Table~\ref{tab:aff_comparison2} compares the performance of CNN-based and LLM-based methods for predicting binding affinity to the same peptide. We used the 14H dataset (heavy chain sequences) and 14L dataset (light chain sequences) which are the mutated version of Ab-14 heavy and light chain sequences, respectively. To fit the heavy chain sequences from 14H dataset, the seed light chain sequence of Ab-14 was paired with different heavy chains from 14H to form scFv sequences. Similarly, the seed heavy chain was paired with different light chains from 14L to form additional scFv sequences. We observed that the Pearson correlation coefficient of both 14H and 14L antibodies generated Ab-Affinity were the highest among all models. While the best Spearman correlation for 14H was obtained by A2Binder, that of Ab-Affinity closely followed. Note that these correlation coefficients are specific to predicting the binding affinity for antibodies derived from the Ab-14 seed antibody. However, Ab-Affinity not only shows a superior correlation for Ab-14 but also for all three seed antibodies ( Figure~\ref{fig:aff_comp}).


\subsection{Ab-Affinity's Embeddings enable Downstream Classification Tasks}

\begin{figure}[h]
    \centering
    \includegraphics[width=\linewidth]{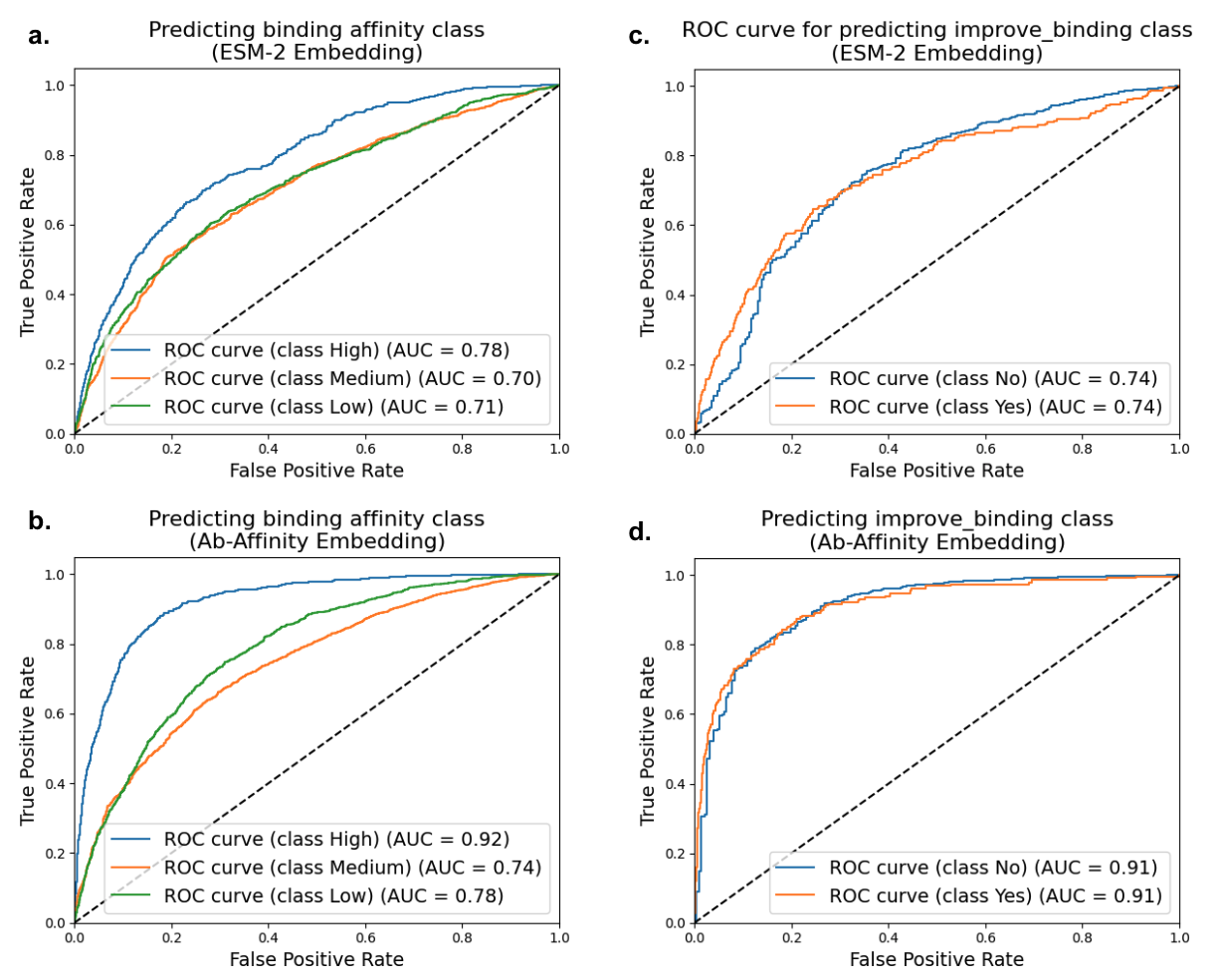} 
    \caption{ROC curves and AUC values for two classification tasks, namely \textbf{(a,b)} determining the binding affinity class of an antibody (High, Medium, Low) and \textbf{(c,d)} determining whether the binding is improved compared to the seed antibody (Y/N); \textbf{(a,c)} ROC curves using the ESM-2 embedding; \textbf{(b,d)} ROC curves using the Ab-Affinity embedding; AUC values are reported for each classifier}
    \label{fig:classification_1}
\end{figure}

We used Ab-Affinity embeddings to carry out two downstream classification tasks related to antibody binding characteristics, namely (i) the problem of determining the antibody binding affinity classes (High, Medium, Low), and (ii) the problem of determining whether the binding affinity of an antibody was stronger than the binding affinity of the corresponding seed antibody (Yes, No). Figure~\ref{fig:classification_1} reports the Receiver Operating Characteristic (ROC) curves for the two classification tasks, comparing the classifiers' AUC using the embeddings produced by Ab-Affinity against ESM-2. For both tasks the classifiers built using Ab-Affinity's embedding has much higher AUC values, suggesting that Ab-Affinity's embeddings are more informative for downstream applications.

\begin{figure}[h]
    \centering
    \includegraphics[width=1.\linewidth]{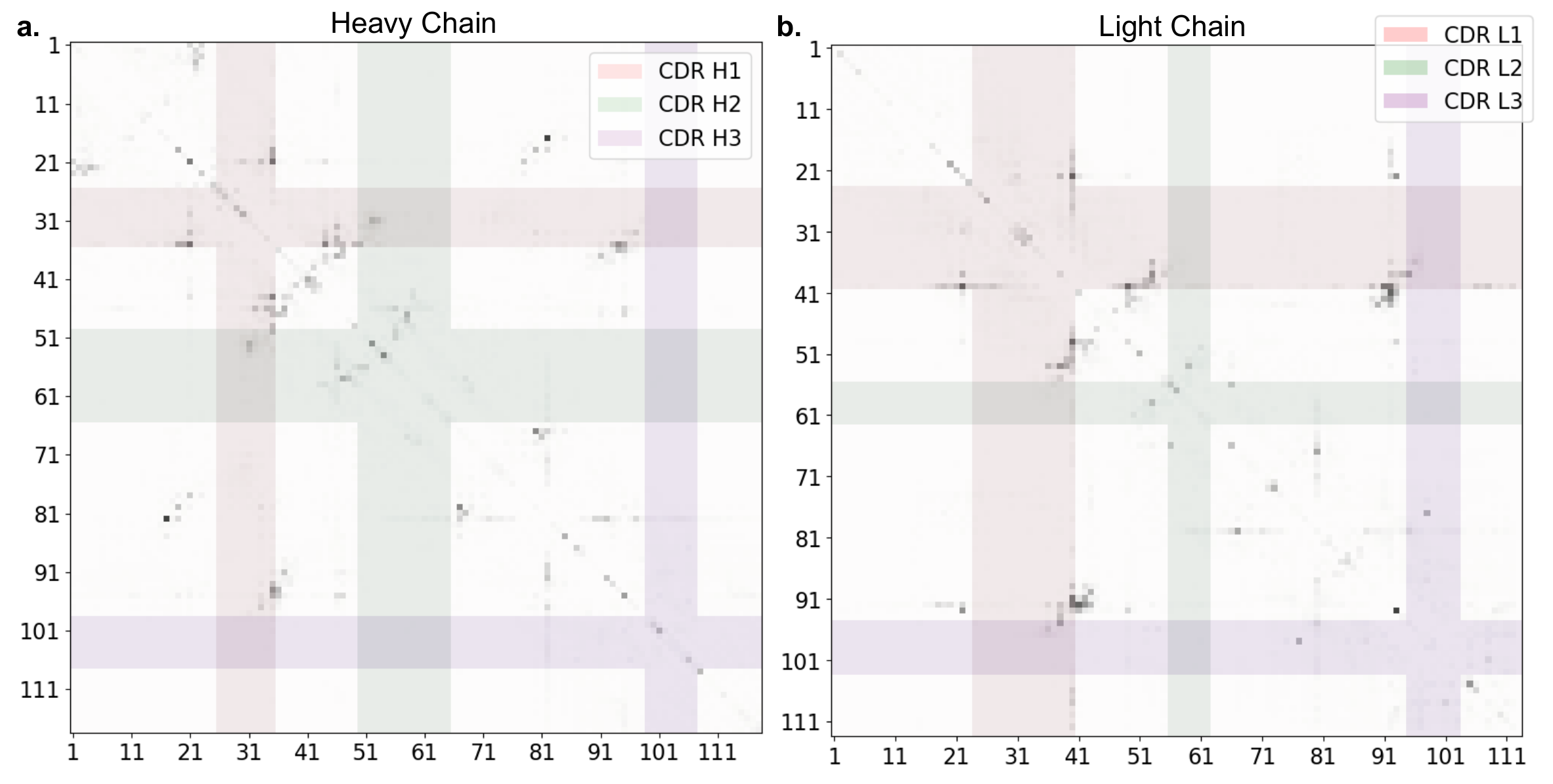} 
    \caption{Differences between the attention residue-residue maps for strong binding (i.e. $\log K_d < 0.5$) and weak binding (i.e., $\log K_d > 5.5$)  for antibodies generated from Ab-14 (a) heavy chain, and (b) light chain.}
    \label{fig:contacts}
\end{figure}

\subsection{Ab-Affinity's Attention Maps for Strong and Weak Binding Antibodies}

We extracted attention maps from the Ab-Affinity model to determine which residue-residue interactions were more important for the prediction of binding affinity. We produced residue-residue contact maps for heavy and light chains of antibodies with strong binding affinity (i.e., $\log K_d < 0.5$) and antibodies with weak binding affinity (i.e., $\log K_d> 5.5$). Figure~\ref{fig:contacts} illustrates the residue-residue differences between the attention maps for strong and weak binding antibodies for both heavy and light chains. The CDRs are highlighted in pink, green and purple in Figure~\ref{fig:contacts}. Observe that many of the strongest differences occur in CDR-H1, CDR-H2, CDR-L1 or the regions immediately adjacent to them.

\subsection{Ab-Affinity Captures Thermostabilty Properties of Antibodies}

\begin{figure}[h]
    \centering
    \includegraphics[width=\linewidth]{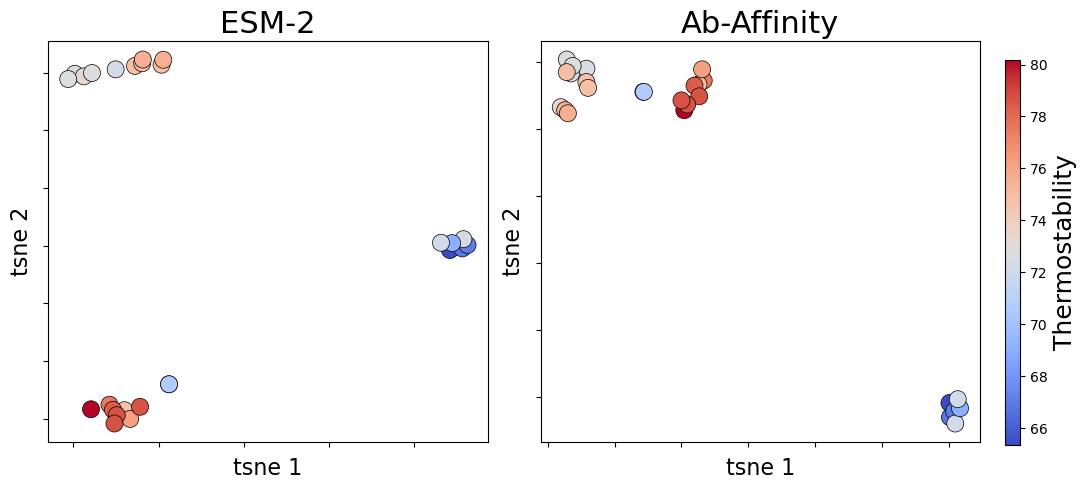}
    \caption{t-SNE visualization of the embeddings produced by ESM-2 and Ab-Affinity; points are colored according to the experimentally-determined thermostability of those antibodies}
    \label{fig:thermostability}
\end{figure}

The ability of a protein to resist degradation or changes to its physical structure at higher temperature is called \emph{thermostability}. This property is essential for therapeutic antibodies as it determines their structural integrity and functionality at higher temperature. Here we investigated the correlation of Ab-Affinity's embeddings with the thermostability of the corresponding antibody. For this task, we used studies \cite{hie2024efficient,rosace2023automated} to create a small dataset of experimentally-determined thermostability for 26 SARS-CoV-2 antibodies.  In the t-SNE plots in Figure~\ref{fig:thermostability} each point is an antibody and the color illustrates its corresponding thermostability. Observe that in the Ab-Affinity's embeddings, antibodies are clearly separated into two clusters, each with relatively similar thermostability values. It is not the case for the ESM-2 embeddings. One explanation of this finding is that the fine-tuned language model has learned intrinsic rules that not only govern the physics of antigen-antibody interaction but also the physics of protein stability that contributes to the stable interactions over a range of temperatures. 

\section{Conclusion}

We introduced Ab-Affinity, a large language model that can predict the binding affinity of antibodies against the SARS-CoV-2 spike protein. In our experiments, Ab-Affinity demonstrated stronger predictive performance than other methods in the literature. The model was able to learn the effect of mutations on binding with the SARS-CoV-2 spike protein. The t-SNE visualization of Ab-Affinity's embeddings suggested that our approach captured the relationships between antibody sequences and binding affinities. Our experiments also show that Ab-Affinity's embeddings enabled a strong performance on several classification tasks, emphasizing the utility of our model beyond affinity prediction.  The residue-residue attention maps extracted from Ab-Affinity reveal that our model appears to focus on the CDR or the neighboring regions, which are the binding sites in the antibody. Ab-Affinity also demonstrated an ``understanding'' of thermostability, despite not being explicitly trained on this attribute. 

In conclusion, Ab-Affinity represents a significant advancement in the prediction of antibody binding affinities, demonstrating both high accuracy and versatility in its applications. Our findings highlight its potential as a powerful tool for antibody design and further exploration of protein-antibody interactions. 

The model is readily available for use and can be easily installed on any machine via PyPi. For convenience, users can access and install Ab-Affinity directly from the Python Package Index (PyPi) using the following link: https://pypi.org/project/AbAffinity/. This ensures seamless integration into various research environments and workflows. 

\section{Acknowledgment} 
The work reported here was funded by the National Institute of Allergy and Infectious Diseases (NIAID) of the National Institutes of Health (NIH ) under the grant 3R01AI169543 to SL and AR.


\begin{quote}
\begin{small}
\bibliographystyle{aaai}
\bibliography{references}

@article{macraild2016antibody,
  title={Antibody recognition of disordered antigens},
  author={MacRaild, Christopher A and Richards, Jack S and Anders, Robin F and Norton, Raymond S},
  journal={Structure},
  volume={24},
  number={1},
  pages={148--157},
  year={2016},
  publisher={Elsevier}
}

@article{uversky2021mobility,
  title={Mobility and disorder in antibody and antigen binding sites do not prevent immunochemical recognition},
  author={Uversky, Vladimir N and Van Regenmortel, Marc HV},
  journal={Critical Reviews in Biochemistry and Molecular Biology},
  volume={56},
  number={2},
  pages={149--156},
  year={2021},
  publisher={Taylor \& Francis}
}

@article{engelhart2022dataset,
  title={A dataset comprised of binding interactions for 104,972 antibodies against a SARS-CoV-2 peptide},
  author={Engelhart, Emily and Emerson, Ryan and Shing, Leslie and Lennartz, Chelsea and Guion, Daniel and Kelley, Mary and Lin, Charles and Lopez, Randolph and Younger, David and Walsh, Matthew E},
  journal={Scientific Data},
  volume={9},
  number={1},
  pages={653},
  year={2022},
  publisher={Nature Publishing Group UK London}
}

@article{devlin2018bert,
  title={Bert: Pre-training of deep bidirectional transformers for language understanding},
  author={Devlin, Jacob and Chang, Ming-Wei and Lee, Kenton and Toutanova, Kristina},
  journal={arXiv preprint arXiv:1810.04805},
  year={2018}
}

@article{lin2023evolutionary,
  title={Evolutionary-scale prediction of atomic-level protein structure with a language model},
  author={Lin, Zeming and Akin, Halil and Rao, Roshan and Hie, Brian and Zhu, Zhongkai and Lu, Wenting and Smetanin, Nikita and Verkuil, Robert and Kabeli, Ori and Shmueli, Yaniv and others},
  journal={Science},
  volume={379},
  number={6637},
  pages={1123--1130},
  year={2023},
  publisher={American Association for the Advancement of Science}
}

@article{vaswani2017attention,
  title={Attention is all you need},
  author={Vaswani, Ashish and Shazeer, Noam and Parmar, Niki and Uszkoreit, Jakob and Jones, Llion and Gomez, Aidan N and Kaiser, {\L}ukasz and Polosukhin, Illia},
  journal={Advances in neural information processing systems},
  volume={30},
  year={2017}
}

@article{van2008visualizing,
  title={Visualizing data using t-SNE.},
  author={Van der Maaten, Laurens and Hinton, Geoffrey},
  journal={Journal of machine learning research},
  volume={9},
  number={11},
  year={2008}
}

@article{hie2024efficient,
  title={Efficient evolution of human antibodies from general protein language models},
  author={Hie, Brian L and Shanker, Varun R and Xu, Duo and Bruun, Theodora UJ and Weidenbacher, Payton A and Tang, Shaogeng and Wu, Wesley and Pak, John E and Kim, Peter S},
  journal={Nature Biotechnology},
  volume={42},
  number={2},
  pages={275--283},
  year={2024},
  publisher={Nature Publishing Group US New York}
}

@article{rosace2023automated,
  title={Automated optimisation of solubility and conformational stability of antibodies and proteins},
  author={Rosace, Angelo and Bennett, Anja and Oeller, Marc and Mortensen, Mie M and Sakhnini, Laila and Lorenzen, Nikolai and Poulsen, Christian and Sormanni, Pietro},
  journal={Nature communications},
  volume={14},
  number={1},
  pages={1937},
  year={2023},
  publisher={Nature Publishing Group UK London}
}

@article{singh2023learning,
  title={Learning the language of antibody hypervariability},
  author={Singh, Rohit and Im, Chiho and Qiu, Yu and Mackness, Brian and Gupta, Abhinav and Sorenson, Taylor and Sledzieski, Samuel and Erlach, Lena and Wendt, Maria and Nanfack, Yves Fomekong and others},
  journal={bioRxiv},
  pages={2023--04},
  year={2023},
  publisher={Cold Spring Harbor Laboratory}
}

@article{barton2024enhancing,
  title={Enhancing antibody language models with structural information},
  author={Barton, Justin and Galson, Jacob D and Leem, Jinwoo},
  journal={bioRxiv},
  pages={2023--12},
  year={2024},
  publisher={Cold Spring Harbor Laboratory}
}

@article{liu2020antibody,
  title={Antibody complementarity determining region design using high-capacity machine learning},
  author={Liu, Ge and Zeng, Haoyang and Mueller, Jonas and Carter, Brandon and Wang, Ziheng and Schilz, Jonas and Horny, Geraldine and Birnbaum, Michael E and Ewert, Stefan and Gifford, David K},
  journal={Bioinformatics},
  volume={36},
  number={7},
  pages={2126--2133},
  year={2020},
  publisher={Oxford University Press}
}

@article{he2024novo,
  title={De novo generation of SARS-CoV-2 antibody CDRH3 with a pre-trained generative large language model},
  author={He, Haohuai and He, Bing and Guan, Lei and Zhao, Yu and Jiang, Feng and Chen, Guanxing and Zhu, Qingge and Chen, Calvin Yu-Chian and Li, Ting and Yao, Jianhua},
  journal={Nature Communications},
  volume={15},
  number={1},
  pages={6867},
  year={2024},
  publisher={Nature Publishing Group UK London}
}

@article{barton2024generative,
  title={A generative foundation model for antibody sequence understanding},
  author={Barton, Justin and Gaspariunas, Aretas and Yadin, David A and Dias, Jorge and Nice, Francesca L and Minns, Danielle H and Snudden, Olivia and Povall, Chelsea and Valle Tomas, Sara and Dobson, Harry and others},
  journal={bioRxiv},
  pages={2024--05},
  year={2024},
  publisher={Cold Spring Harbor Laboratory}
}

@article{shen2024tag,
  title={Tag-LLM: Repurposing General-Purpose LLMs for Specialized Domains},
  author={Shen, Junhong and Tenenholtz, Neil and Hall, James Brian and Alvarez-Melis, David and Fusi, Nicolo},
  journal={arXiv preprint arXiv:2402.05140},
  year={2024}
}

@article{myung2022csm,
  title={CSM-AB: graph-based antibody--antigen binding affinity prediction and docking scoring function},
  author={Myung, Yoochan and Pires, Douglas EV and Ascher, David B},
  journal={Bioinformatics},
  volume={38},
  number={4},
  pages={1141--1143},
  year={2022},
  publisher={Oxford University Press}
}

@article{yuan2023dg,
  title={DG-Affinity: predicting antigen--antibody affinity with language models from sequences},
  author={Yuan, Ye and Chen, Qushuo and Mao, Jun and Li, Guipeng and Pan, Xiaoyong},
  journal={BMC bioinformatics},
  volume={24},
  number={1},
  pages={430},
  year={2023},
  publisher={Springer}
}

@article{ozturk2018deepdta,
  title={DeepDTA: deep drug--target binding affinity prediction},
  author={{\"O}zt{\"u}rk, Hakime and {\"O}zg{\"u}r, Arzucan and Ozkirimli, Elif},
  journal={Bioinformatics},
  volume={34},
  number={17},
  pages={i821--i829},
  year={2018},
  publisher={Oxford University Press}
}

@article{abbasi2020island,
  title={ISLAND: in-silico proteins binding affinity prediction using sequence information},
  author={Abbasi, Wajid Arshad and Yaseen, Adiba and Hassan, Fahad Ul and Andleeb, Saiqa and Minhas, Fayyaz Ul Amir Afsar},
  journal={BioData Mining},
  volume={13},
  pages={1--13},
  year={2020},
  publisher={Springer}
}

@article{li2023machine,
  title={Machine learning optimization of candidate antibody yields highly diverse sub-nanomolar affinity antibody libraries},
  author={Li, Lin and Gupta, Esther and Spaeth, John and Shing, Leslie and Jaimes, Rafael and Engelhart, Emily and Lopez, Randolph and Caceres, Rajmonda S and Bepler, Tristan and Walsh, Matthew E},
  journal={Nature Communications},
  volume={14},
  number={1},
  pages={3454},
  year={2023},
  publisher={Nature Publishing Group UK London}
}

@article{rao2020transformer,
  author = {Rao, Roshan M and Meier, Joshua and Sercu, Tom and Ovchinnikov, Sergey and Rives, Alexander},
  title={Transformer protein language models are unsupervised structure learners},
  year={2020},
  doi={10.1101/2020.12.15.422761},
  url={https://www.biorxiv.org/content/10.1101/2020.12.15.422761v1},
  journal={bioRxiv}
}
\end{small}
\end{quote}

\end{document}